\documentclass[aps]{revtex4-1}
\usepackage{amsmath,amssymb}
\usepackage{graphicx}
\usepackage{dcolumn}
\usepackage{bm}
\usepackage{color}

\begin{document}
\title{Temperature-dependent collective effects for silicene, germanene and monolayer black phosphorus}

\author{Andrii Iurov}
\email{aiurov@unm.edu}
\affiliation{
Center for High Technology Materials, University of New Mexico,
1313 Goddard SE, Albuquerque, NM, 87106, USA
}
\author{Godfrey Gumbs}
\affiliation{
Department of Physics and Astronomy, Hunter College of the City
University of New York, 695 Park Avenue, New York, NY 10065, USA
}
\affiliation{
Donostia International Physics Center (DIPC),
P de Manuel Lardizabal, 4, 20018 San Sebastian, Basque Country, Spain
}
\author{Danhong Huang}
\affiliation{
Air Force Research Laboratory, Space Vehicles Directorate,
Kirtland Air Force Base, NM 87117, USA
}
\affiliation{
	Center for High Technology Materials, University of New Mexico,
	1313 Goddard SE, Albuquerque, NM, 87106, USA
}

\date{\today}

\begin{abstract}
We have calculated numerically electron exchange, correlation energies and dynamical
polarization function for newly discovered silicene, germanene and black phosphorus (BP),
consisting of  puckered layers of elemental phosphorus atoms, broadening the
range of two-dimensional (2D) materials  at various temperatures. As a  matter of fact,
monolayer BP, produced by mechanical and liquid exfoliation techniques,
has been predicted to be an insulator with a large  energy splitting
$\approx 1.6\,$eV for the quasiparticle band structure.
We compare  the dependence  of  these energies  on
the chemical potential, field-induced gap and the temperature  and concluded that in
many cases they behave qualitatively similarly, i.e., increasing with the
doping, decreasing significantly at elevated temperatures, and displaying different
dependence on the asymmetry gap at various temperatures. Furthermore, we used the
dynamical polarizability to investigate the new ``split'' plasmon branches in the puckered
lattices and predict a unique splitting, different from that in gapped graphene,
for various energy gaps. Our results are crucial for stimulating electronic, transport
and collective studies of these novel materials,  as well as for enhancing silicene-based
fabrication and technologies for photovoltaics and transistor devices.
\end{abstract}
\maketitle

Since high-quality graphene crystal became available in 2004, its  transport and optical  properties
due to quantum confinement have been extensively studied both theoretically  and experimentally. In
particular,  free-standing graphene is a semiconductor  but   silicene,  a  buckled structure  in
which the silicon atoms are displaced perpendicular to the basal plane, as  well as black phosphorus
have been broadening interest in two-dimensional materials because of the presence of a larger gap
between the quasiparticle energy bands. This gives rise to the possibility of novel
practical applications.\,\cite{gr1,gr2,gr3} It is well known that graphene exhibits a number of
interesting properties which are  related  to its novel electronic structure near the Fermi level,
represented by the so-called Dirac cone. Due to this unique linear dispersion and chirality, graphene
electrons may not be confined by electrostatic gate voltage or energy barriers. Such a situation
changes when there is  a gap between the valence and conduction bands. Consequently,
gap opening and its related tunability become a crucial issue for developing novel transistor technologies.
This goal could be achieved by employing graphene nanoribbons, putting graphene on
various insulating substrates\,\cite{gap1, gap2, gap3, gap4} or even under illumination  by circularly-polarized light.\,\cite{kibis}
\medskip
\par

Flexibly-tuned phase transitions between insulating and conducting states were   predicted
theoretically and then demonstrated experimentally  for  puckered  lattices of other fourth- and fifth-group
elements, such as silicene, germanene and black phosphorus \cite{BP1,BP2}. In contrast to the carbon-based graphene,
Si and Ge atoms have a larger ionic radius, making sp$^3$ hybridization energetically favorable
and leading to out-of-plane buckling and asymmetry between sublattices. Obviously, the
bandgap induced by buckling could be tuned by applying an electric field perpendicular
to the silicene plane. Such predicted bandgap tunability may be easily realized
in experiments.\,\cite{SilMain, ezawa, ezawaprl}
\medskip

Following the epitaxial growth of silicene in 2010, considerable attention
has been devoted  to understanding  its microscopic electronic properties which have led
to silicene being an   excellent material candidate due to its significant spin-orbit coupling
(SOC) and electrically-tunable properties. A key effect of an applied electric field
is that it can either open or close the energy band gap, which is an important requirement
for digital electronics applications. A variety    of fascinating features of silicene
have been  discussed from a theoretical point of view. These include 	the
quantum spin Hall and anomalous Hall  effect,   as well as valley-polarized
quantum Hall effect under an electric field, anomalous Hall insulators and single-valley semimetals, potential giant magneto-resistance, and topologically protected
helical edge states.\,\cite{Dru4,Dru5,Dru6}  Germanene, on the other hand, is believed
to acquire a much larger intrinsic spin-orbit gap $24$-$93\,$meV\,\cite{G1zhang,G2acun},
and only very few  groups have managed to synthesize
germanene\,\cite{G11li, G12davila, G13bampoulis, G14derivaz} on different substrates.
\medskip

A careful theoretical treatment of electron correlations in the materials
under investigation  is necessary since  a qualitative difference between
the results may be obtained from the employment  of specific methods for calculations. The method we have used
might be considered as a starting point for further analysis and may
trigger higher level approximations in future works. Analytical  calculation of
the correlation energy requires a screened Coulomb interaction within
the random-phase approximation (RPA). Earlier   calculations of exchange and correlation energies have been
reported for  conventional 2D and  3D   electron gases.\,\cite{harris,harris2}
It was pointed out by Barlas\,\cite{barlas, polini2} that both charge and spin
susceptibilities in graphene are significantly reduced due to electron chirality,
and this reduction is further enhanced at lower electron densities.
Therefore, for silicene it is very interesting to see how the exchange and
correlation energies are modified in the presence of a gap but the absence of electron
chirality at the same time.
\medskip

\begin{figure}
\centering
\includegraphics[width=0.55\textwidth]{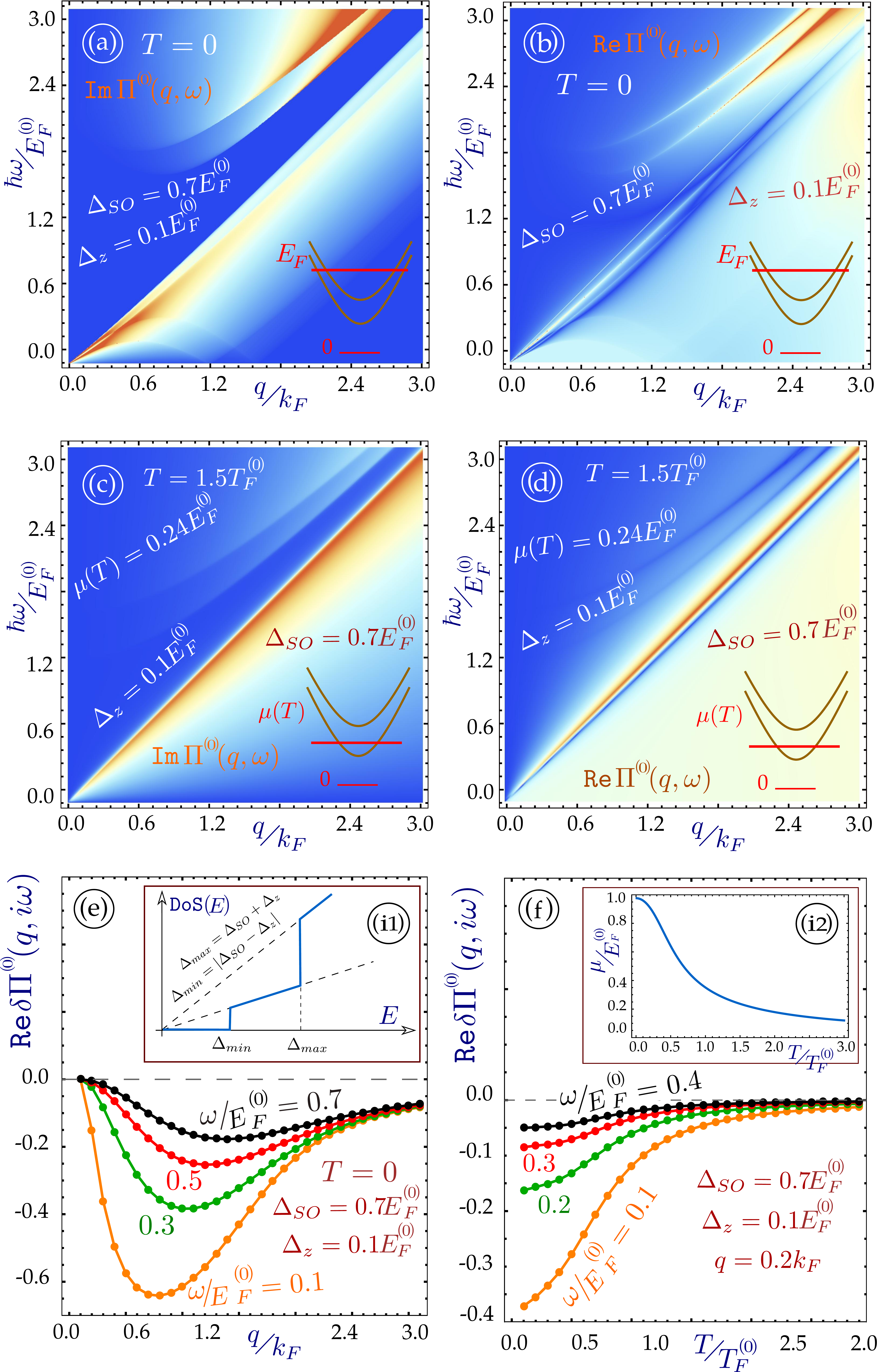}
\caption{(Color online)  Panels $(a)$ and $(b)$  present density plots of the 
imaginary and real parts of the non-interacting  polarizability 
$\Pi^{(0)}(q,\omega)$ at $T=0$ as functions of frequency
in units of $\mu(T=0)=E_F^{(0)}$  for
$\Delta_{\rm SO}=0.7\, E_F^{(0)}$ and $\Delta_{\rm z}=0.1\, E_F^{(0)}$ (both spin-dependent subbands
are doped and have energy gaps). Plots $(c)$ and $(d)$ show similar results at $T=T^{(0)}_F\equiv E_F^{(0)}/k_B$.
Panel $(e)$ displays the real part of the shifted polarizability  $\delta \Pi^{(0)}(q,i\omega)$ 
due to doping, as defined in Eq.\ (\ref{eqn5}),  as a function of $q$ at $T=0$
for various $i\omega$ values, while plot $(f)$ exhibits the real part of $\delta\Pi^{(0)}(q,\omega)$
as a function of $T$ for  $q=0.2\,k_F$ ($k_F=E^{(0)}_F/\hbar v_F$) and various $i\omega$ values.
All the results are scaled by the graphene density of states
$\mbox{DoS}(E^{(0)}_F)=2 E^{(0)}_F/ \pi\hbar^2v_F^2$ at the Fermi energy. Inset
$(i1)$ presents $\mbox{DoS}(E)$ of silicene, while inset $(i2)$
corresponds to $\mu(T)$ of doped silicene.}
\label{FIG:1}
\end{figure}

The low-energy model Hamiltonian for a puckered honeycomb lattice could be written as
\begin{equation}
{\cal H}_\xi = \hbar v_F (\xi k_x \hat{\tau}_x + k_y \hat{\tau}_y)
- \xi \Delta_{\rm SO} \hat{\sigma}_z \hat{\tau}_z + \Delta_z \hat{\tau}_z\ .
\label{eqn1}
\end{equation}
Here, the first term is the graphene Hamiltonian with $v_F=5\times 10^{5}\,$m/s,
$\hat{\tau}_{x,y,z}$ and $\hat{\sigma}_{x,y,z}$ are Pauli matrices corresponding to two sub-spaces,
$\xi = \pm 1$ for $K$ and $K^\prime$ points. The second term describes a
Kane-Mele system\,\cite{KaneMele} with a spin-obrit gap $ \Delta_{\rm SO} $.
The last term represents the sublattice potential difference with $\Delta_z = E_z d$
due to a perpendicular electric field $E_z$, where $d \backsimeq
0.46\,$\AA\ is the out-of-plane sublattice displacement due to buckling.
\medskip

The Hamiltonian in Eq.\,(\ref{FIG:1}) yields the low-energy dispersion relations
\begin{equation}
s E_k = \pm \sqrt{(\hbar v_F)^2 k^2 + \Delta_{\sigma,\xi}^2}\ ,
\label{eqn2}
\end{equation}
where $sE_k$ is the energy dispersion with $s=1$ ($s=-1$) for the conduction (valence) band,
$\Delta_{\sigma,\xi} = \vert \sigma \xi \Delta_{\rm SO} - \Delta_z \vert$ with $\sigma = \pm 1$
represents two spin-resolved energy bandgaps, determined by either the sum
or the difference between two gaps for two spin states
(only electrons near the $K$ point with $\xi=1$ are considered).
Therefore, we can write $\Delta_> = \Delta_{\rm SO}+\Delta_z$ and $\Delta_{<}
= \vert  \Delta_{\rm SO} - \Delta_z \vert$. A similar model Hamiltonian may be used to describe BP
near the $\Gamma$  and $\Gamma^\prime$ points.
\medskip

When $E_z=0$, the asymmetry gap $\Delta_z = 0$ and the system behaves as a $Z_2$
topological  insulator (TI) or  spin-Hall  insulator due to the fact that two edge
states in a silicene nanoribbon become gapless\,\cite{ezawa,ezawa9prl} and
$\Delta_{<}=\Delta_{>}= \Delta_{\rm SO}$. Once $E_z>0$, the minimal gap
$\Delta_{<}$ falls off  linearly to zero, approaching a gapless conducting state.
In this situation, only one energy gap, corresponding to a specific pair of subbands
with $\sigma=1$ is closed. Such a state is referred to as a \textit{valley-spin
polarized metal} (VSPM). If $E_z$is further increased beyond this point, the system
transfers to a trivial band insulator (BI) with no gapless edge states even far away
from the $K$-points, and the energy gap keeps growing and never closes again.
\medskip

The dynamical polarization function $\Pi^{(0)}(q,\omega)$ is one of the most important
quantities to determine the transport, collective and charge screening properties of
a material. In the one-loop approximation, for a clean sample without impurity and
phonon scatterings at low temperatures,   is given by
\begin{widetext}
\begin{equation}
\Pi^{(0)}(q,\omega) = \frac{1}{8 \pi^2} \sum\limits_{\sigma, \xi = \pm 1}
\int d^2 {\bf k} \sum\limits_{s, s' = \pm 1}
\left\{1+ s s'\frac{ \hbar^2v_F^2({ \bf k}+ { \bf q})\cdot { \bf k} +
\Delta_{\sigma,\xi}^2}{ E_k\,E_{|{\bf k+q}|}} \right\}
\frac{f_0(s E_{k}-\mu,T)-f_0(s' E_{|{\bf k+q}|}-\mu,T)}{s E_k - s' E_{|{\bf k+q}|}
-\hbar (\omega + i 0^+)} \ ,
\label{eqn3}
\end{equation}
\end{widetext}
where  $f_0(s E_{k}-\mu,T)$ is the Fermi-Dirac distribution function for a given subband,
$\mu$ and $T$ are the chemical potential and temperature of the system.
Equation\ (\ref{eqn3}) is just an average of two graphene polarizabilities with two
inequivalent energy gaps $\Delta_{<}$ and $\Delta_{>}$, i.e., two subband pairs contribute
equivalently. Such a fact is often used in describing many-body properties of
silicene.\,\cite{SilMain} The imaginary part of $\Pi^{(0)}(q,\omega)$,
presented in Fig.\ref{FIG:1} $(a)$ and $(c)$ shows the single-particle excitation
regions or the particle-hole modes (PHM's), i.e., those regions
where a plasmon is Landau damped by breaking it into  electron-hole excitations.
\medskip

\begin{figure}
\centering
\includegraphics[width=0.55\textwidth]{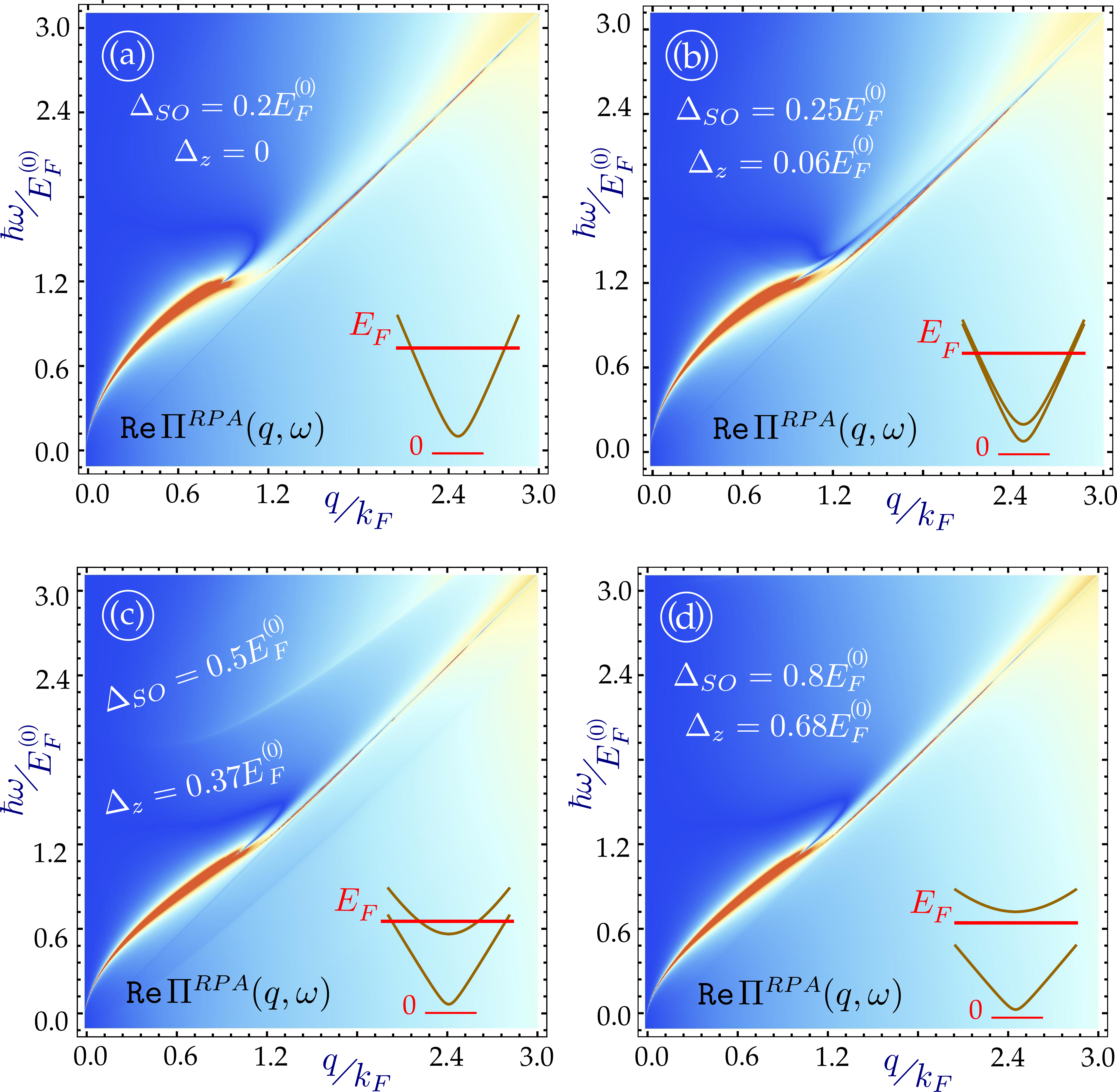}
\caption{(Color online)  Split plasmon branches (peaks of the RPA polarization function) for
different combinations of energy gaps $\Delta_{\rm SO}$
and $\Delta_z$ in panels $(a)$, $(b)$ and $(c)$. Panel $(d)$ corresponds to
the case with an unfilled upper subband.}
\label{FIG:2}
\end{figure}

Plasmons are obtained from zero of the dielectric function given by
$\epsilon^{(0)}(q,\omega) = 1 -V(q)\,\Pi^{(0)}(q,\omega)$, where $V(q)=e^2/2\epsilon_0\epsilon_bq$ 
is the two-dimensional Coulomb interaction and $\epsilon_b$ is the dielectric constant of the host material.
Examples of plasmons in silicene are shown in Fig.\,\ref{FIG:2} and will be discussed below.
\medskip

The central subject of the current study is the $T$ dependence of the many-body
effects in silicene. It is known that $\mu$ decreases with $T$ for fixed doping level.
The existence of an infinite sea of holes below the Dirac points guarantees $\mu>0$
and $\mu\rightarrow 0$ in the high-$T$ limit. In order to obtain the result for
silicene, we have to calculate the density of states. Our early study\,\cite{myT} revealed
that the density of states of graphene is not changed in the presence of an energy
bandgap for the energies above the gap. This should also be valid for
silicene with the two pairs of subbands separated by energy gaps. The calculated
density of states $\mbox{DoS}(E)$ is a piece-wise linear function, which is presented
in the inset $(i1)$ of Fig.\,\ref{FIG:1}, and the chemical potential $\mu(T)$ for
silicene, quite similar to graphene, is shown in another inset $(i2)$.
\medskip

\begin{figure}
\centering
\includegraphics[width=0.55\textwidth]{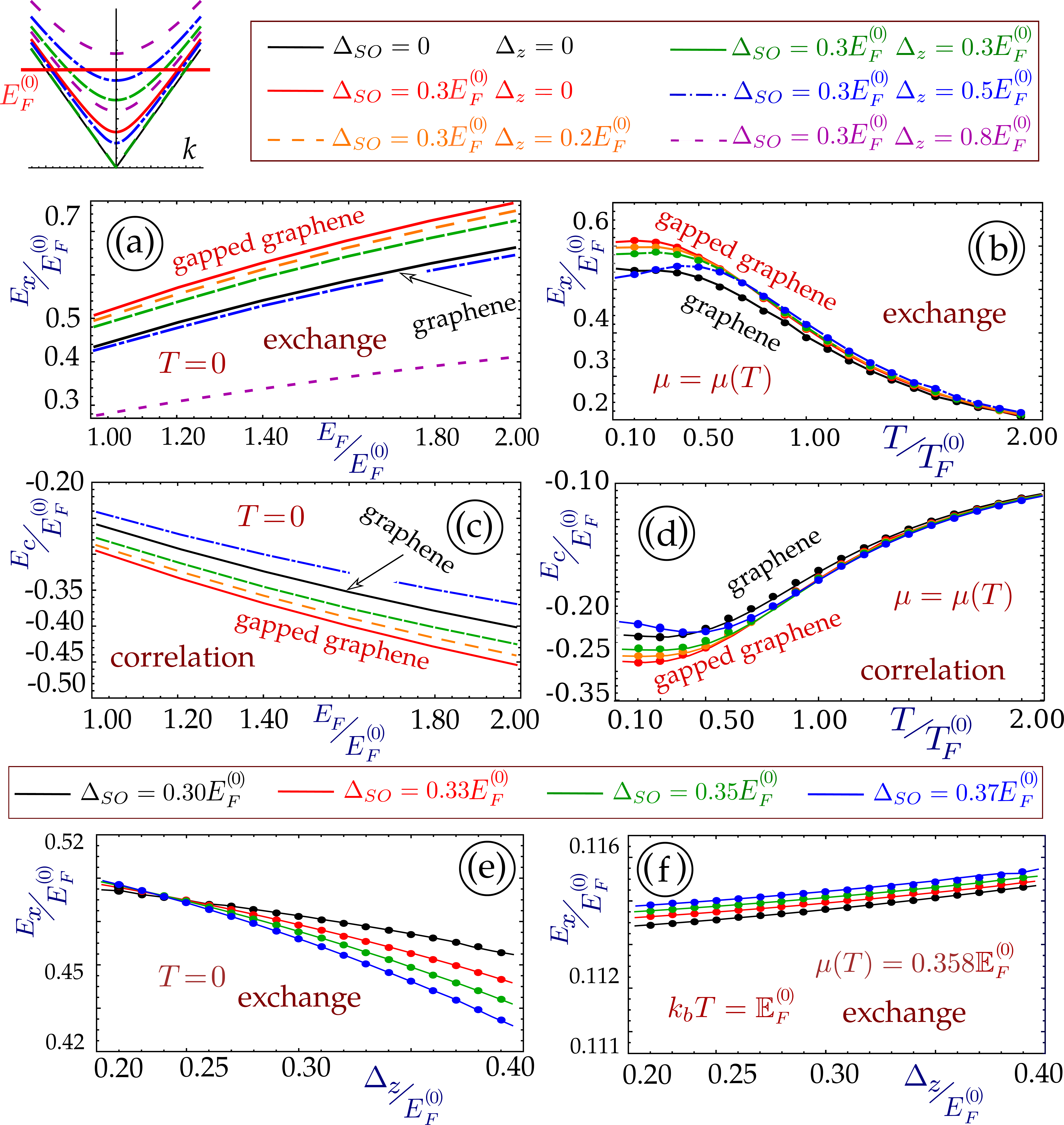}
\caption{(Color online)  Exchange ($E_x$) and correlation ($E_c$) energies for graphene and silicene with different energy gaps at various $T$.
Plots $(a)$ and $(c)$ demonstrate doping dependence of $E_x$ and $E_c$ at $T=0$ for graphene
(black solid curve), gapped graphene or silicene with $\Delta_z=0$ (red solid curve) and for different combinations of $\Delta_{\rm SO}$ and $\Delta_z$.
Panels $(b)$ and $(d)$ show the $T$ dependence of $E_x$ and $E_c$ with fixed doping
($E_F \equiv \mu(T=0) = E_F^{(0)}$). Plots $(e)$ ($T=0$) and $(f)$ ($T=E_F/k_B$) display the $\Delta_z$ dependence of silicene $E_x$
with different $\Delta_{\rm SO}$ to demonstrate the TI$\rightarrow$VSPM$\rightarrow$BI phase transition.}
\label{FIG:3}
\end{figure}

Although plasmon excitations in silicene have attracted some attentions recently,\,\cite{SilMain, plas2, Wu}
there is an interesting experimental observation which is still lack of a clear
explanation. For a graphene with an energy gap in the range of
$\Delta_0 \backsimeq (0.20 - 0.22)\,\mu$, the plasmon mode was found first to enter into
and then exit from the particle-hole continuum, leading to \textit{split or ``broken''}
undamped plasmon  branch.\,\cite{pavlo1} Such an extended undapmed plasmon mode is of key
interest to a number of applications, such as plasmonic metamaterials, nanoarrays and
tunable hybrid optical devices.\,\cite{polinano} For silicene, we have found this split
plasmon occurs within a compensated spin-orbit gap $\Delta_{\rm SO}$ by a field-induced one $\Delta_z$. Our numerical results for a split plasmon branch with
various combinations of these two gaps are shown in Fig.\,\ref{FIG:2}.
The minimum gap $\Delta_{<}$ determines the outermost boundary of the PHM region and gives rise to
a PHM-free region for the plasmon branch at finite wave vector $q$. In general, each of two gaps
affects the behavior of the plasmon branch independently. If both spin-split subbands are
occupied as $\mu > \Delta_{>}$, the plasmon frequency in the long-wavelength limit becomes
$\backsimeq \sqrt{(\mu^2 - \Delta_{\rm SO}^2-\Delta_z^2)\,q}$. In this case, both gaps play
the same role by reducing the plasmon frequency. Although we cannot derive an analytic
expression for the plasmon dispersion relation at large $q$ values, the general feature from our
numerical results in Fig.\,\ref{FIG:2} may be summarized as follows: both $\Delta_{\rm SO}$
and  $\Delta_z$ reduce the plasmon frequency, causing the plasmon branch to stay closer to the
main diagonal. We further note that the split plasmon branch survives even for
the case with only one occupied subband, as shown in Fig.\,\ref{FIG:2}$(d)$. Here, all
our calculations for the plasmon dispersion relation were conducted at $T=0$.
\medskip

We now turn our attention to calculating the exchange and correlation energies of
silicene. For convenience, we omit the theoretical derivation but simply write down
expressions used in our calculations, based on  the \textit{interaction energy}
integral\,\cite{Gbook} for the jellium model as well as the
fluctuation-dissipation-theorem.\,\cite{barlas, Gui}
Within this model, the calculated energy appears to be divergent. This
divergence, however, can be removed, if we consider only doped
electrons above the Dirac point, i.e. choosing the total energy of intrinsic
graphene as the zero-energy point.\,\cite{barlas, alireza}.
We also apply an upper limit $Q_c$ for the $q$-integral since the Dirac-cone
approximation for silicene is valid only up to $0.46\,$eV.
The calculated exchange energy is
\begin{equation}
E_x = - \frac{\hbar}{2 \pi n} \int\limits_{0}^{Q_C}
\frac{d^2{\bf q}}{(2 \pi)^2} V(q) \int\limits_{0}^{\infty} d\omega\, \delta \Pi^{(0)} (q,i \omega)\ ,
\label{eqn4}
\end{equation}
where the shifted polarizabity related to doping is
\begin{equation}
\delta \Pi^{(0)} (q,i \omega) = \Pi^{(0)} (q,i \omega; \mu) -
\Pi^{(0)} (q,i \omega; \mu = 0) \ .
\label{eqn5}
\end{equation}
At $T=0$, the doping contribution to the polarization function could be easily
separated\,\cite{pavlo1} since the  Fermi-Dirac distribution is just a unit-step
function. The correlation energy is now given in the
RPA by
\[
E_c = -E_x 
\]
\begin{equation}
+ \frac{\hbar}{2 \pi n} \int\limits_{0}^{Q_C} \frac{d^2{\bf q}}{(2 \pi)^2} \int\limits_{0}^{\infty} d\omega\, \ln\left[ \frac{1 - V(q)\,\Pi^{(0)} (q,i \omega; \mu)}{1 - V(q)\,\Pi^{(0)} (q,i \omega; 0)} \right]\ .
\label{eqn6}
\end{equation}

Our numerical results for the exchange-correlation energy  are presented in Fig.\,\ref{FIG:3}.
If silicene is undoped at $T=0$, $\mu(T)=0$ is maintained  for all $T$,
which yields a common reference point.  Results for
$\delta\Pi^{(0)} (q,i \omega)$ with doping as functions of $q$ and $T$ are displayed
in Fig.\,\ref{FIG:1}[$(e)$,$(f)$]. At $T=0$, the doping effect is only limited
to a region covered by small values of $q$ and $\omega$.  In addition,
$\delta \Pi^{(0)} (q,i \omega)$ decreases with increasing $T$.
Such behaviors are expected by noticing very small ${\rm Re}[\Pi^{(0)} (q,i \omega)]$
for the interband transitions above the main diagonal as shown in Figs.\,\ref{FIG:2}[$(b)$,$(c)$].
We also find that $E_x$ and $E_c$, except for being opposite to each other,
behave similarly in various situations. This implies that the $\log$-term in
Eq.\,(\ref{eqn6}) is finite but insignificant to the correlation. At $T=0$, $E_x$
in Fig.\,\ref{FIG:3}($a$) increases with doping since more electrons may contribute.
Interestingly, even the exchange-correlation energy per electron also increases
with the doping, as shown in Fig.\,\ref{FIG:3}[$(a)$,$(c)$], due to enhanced many-body
effects. Moreover, from  Fig.\,\ref{FIG:3}$(e)$,  we find that the exchange energy is  increased
for a larger energy gap, which agrees with earlier reported results in Ref.\,[\onlinecite{alireza}].
Therefore, we may state that $E_x$ and $\vert E_c \vert$ for silicene will increase with $\Delta_{<}$.
On the other hand, $E_x$ decreases as $T$ is increased, similar to $\mu(T)$.
Quantitatively, crossing of curves for different combinations of $\Delta_{\rm SO}$ and
$\Delta_z$ are found in Fig.\,\ref{FIG:3}[$(b)$,$(d)$].
Results for the $\Delta_z$ dependence of the exchange energy on $\Delta_z$ are compared in Fig.\,\ref{FIG:3}[$(e)$,$(f)$] at zero and a finite temperature, where different points for the
TI$\rightarrow$VSPM$\rightarrow$BI transition are found by varying $\Delta_{\rm SO}$.
\medskip

In summary, we have obtained extensive numerical results for the exchange and
correlation energies in silicene at various temperatures and energy gaps
and found that they increase with doping but decrease with temperature.
The $\Delta_z$ dependence of exchange energy with various $\Delta_{\rm SO}$
changes significantly as $T$ increases.
\medskip

In contrast to gapped graphene, for plasmon excitations in silicene, we have
demonstrated the breaking down of a single plasmon branch into two undamped pieces,
appearing for different combinations of $\Delta_{\rm SO}$ and $\Delta_z$. Some of these
combinations correspond to one unoccupied subband with $\mu < \Delta_{>}$.
\medskip

Since silicene, germanine and other puckered  lattices are believed to be
promising replacements for graphene and topological insularors,
we expect that extensive knowledge on plasmonic properies of these materials, such
as exchange and correlation corrections to the electron kinetic energy, local
compressibility and susceptibility, become crucial for applying to fabrication
of semiconductor electronic devices, developing spintronics and many other new areas 
of interests.

\bibliography{SilBib}

\end{document}